# Nondegenerate, Lamb-shift Solution of the Dirac Hydrogen Atom

Nathan S. Roberts

*NASA, Engineering Directorate, Houston, Texas, USA (nathan.s.roberts@nasa.gov)*

**ABSTRACT**. When the Dirac equation was first published in 1928, three solutions appeared immediately within the same year, each describing the most important problem in physics at that time: the hydrogen atom. These solutions lifted some of the degeneracy from earlier atomic models, but not all of it—they still predicted the same degenerate energy levels for the $2s_\frac{1}{2}$ and $2p_\frac{1}{2}$ states, for example. In this paper, we introduce a new solution of the Dirac equation, which finally removes all degeneracy from the hydrogen atom. We work in terms of dimensionless quantities and use the Lamb shift to give each atomic state in our model a unique, nondegenerate energy level. We obtain radial eigenfunctions in terms of the Laguerre polynomials, demonstrate how they can be reduced to the Schrodinger wavefunctions by applying limits, and plot our results.

## 1. INTRODUCTION

In 1928, Paul Dirac introduced a quantum, relativistic equation in four-dimensional (4D) spacetime [1]:

$$(\not{p} - e\not{A} - mc)\psi = 0 \quad (1)$$

The Dirac equation (1) can describe the 4-momentum $p = (p_0, p_1, p_2, p_3)$ of an electron, which is a unit of negative elementary charge $-e$, interacting with an electromagnetic 4-potential $A = (A_0, A_1, A_2, A_3)$. The electron's small mass $m$ enables it to reach speeds near the speed of light $c$, and its spin is defined with a spinor $\psi$ (psi). We use the "sword" metric signature $(+---)$ to expand the slash notation $\not{A} = \gamma^0 A_0 - (\gamma^1 A_1 + \gamma^2 A_2 + \gamma^3 A_3)$, and use boldface type $\not{A} = \gamma^0 A_0 - \boldsymbol{\gamma} \cdot \mathbf{A}$ for spatial components like the vector potential $\mathbf{A} = (A_1, A_2, A_3)$ and momentum $\mathbf{p} = (p_1, p_2, p_3)$ in three-dimensional (3D) space. Timelike gamma matrix $\gamma^0$ and spacelike gamma matrices $\boldsymbol{\gamma} = (\gamma^1, \gamma^2, \gamma^3)$ can be represented by block matrices in the Dirac basis:

$$\gamma^0 = \begin{pmatrix} 1 & 0 \\ 0 & -1 \end{pmatrix} \qquad \boldsymbol{\gamma} = \begin{pmatrix} 0 & \boldsymbol{\sigma} \\ -\boldsymbol{\sigma} & 0 \end{pmatrix} \quad (2)$$

Each matrix entry in (2) is itself a 2×2 submatrix, like the Pauli spin matrices $\boldsymbol{\sigma}$, so (1) can be written as a matrix expression representing two simultaneous equations:

$$\begin{pmatrix} p_0 - eA_0 - mc & -\boldsymbol{\sigma} \cdot (\mathbf{p} - e\mathbf{A}) \\ \boldsymbol{\sigma} \cdot (\mathbf{p} - e\mathbf{A}) & -p_0 + eA_0 - mc \end{pmatrix} \begin{pmatrix} \psi^+ \\ \psi^- \end{pmatrix} = 0 \quad (3)$$

Historically, the "large" spinor components $\psi^+$ link to positive energy solutions, while "small" components $\psi^-$ led to the discovery of antimatter. Dirac replaced the zeroth 4-vector components with energy $p_0 = E/c$ and a central potential $eA_0 = V/c$ [1]. We can represent both equations in one line using plus-minus $\pm$ signs. Thus, multiplying (3) by $\pm c$, then adding $c\boldsymbol{\sigma} \cdot (\mathbf{p} - e\mathbf{A})\psi^\mp$ to both sides gives:

$$(E - V \mp mc^2)\psi^\pm = c\boldsymbol{\sigma} \cdot (\mathbf{p} - e\mathbf{A})\psi^\mp \quad (4)$$

For bound states, we find it convenient to work in terms of the reduced mass $\mu$ (mu), the reduced Planck constant $\hbar$ (h-bar), and the reduced Bohr radius $a_0$ (a-naught):

$$\mu = \frac{1}{\frac{1}{M} + \frac{1}{m}} \qquad \hbar = \frac{h}{2\pi} \qquad a_0 = \frac{\hbar}{Z\alpha c\mu} \quad (5)$$

$M$ is nuclear mass, $h$ is Planck's constant, $2\pi$ is unit circle circumference, $Z$ is atomic number, and $\alpha$ (alpha) is the fine structure constant. Dirac set the vector potential equal to zero ($\mathbf{A} = 0$) and used noncommutative algebra to separate the spinors $\psi^\pm$ into spinor spherical harmonics and radial eigenfunctions $R^\pm$ of radius $r$. In terms of the radial derivative operator $d/dr$, he obtained the form [1]:

$$(E - V \mp \mu c^2)R^\pm = c\hbar \left( \frac{k \mp 1}{r} \mp \frac{d}{dr} \right) R^\mp \quad (6)$$

Due to spin, Dirac realized orbital angular momentum is not a constant of motion, but total angular momentum quantum number $j$ is instead. He also introduced the new quantum number $k$ in (6) that takes up to two values per quantum number $\ell = 0, 1, 2, \ldots, n-1$ (or s, p, d, f, etc.) [1]:

$$k = \pm(j + \tfrac{1}{2}) \quad \text{when} \quad j = \ell \mp \tfrac{1}{2} \quad (7)$$

We can subtract the absolute value of Dirac's quantum number $|k| = j + \tfrac{1}{2}$ from the principal quantum number $n = 1, 2, 3, \ldots$ to obtain a general quantum number:

$$g = n - |k| \quad (8)$$

Three weeks after the Dirac equation (6) was published, Walter Gordon solved it for a hydrogen atom [2]. Just two weeks later, Charles G. Darwin submitted an alternative solution [3]. Fredrick B. Pidduck introduced yet a third solution in 1928 using the Laguerre polynomials [4]. All three solutions predicted the same degenerate energy levels for the $2s_\frac{1}{2}$ and $2p_\frac{1}{2}$ states. Despite the enormous impact of these first three solutions, hardly any new such solutions seem to have appeared after 1928 [5]. In this paper, we introduce a new solution of the Dirac equation (6), which finally removes all degeneracy from the hydrogen atom. Fortunately, in 1947, Willis Lamb and his graduate student Robert Retherford experimentally demonstrated that the $2s_\frac{1}{2}$ energy or frequency level is approximately 1000 MHz (megahertz) higher than the $2p_\frac{1}{2}$ level [6]. One week later, Hans Bethe calculated a nonrelativistic value of 1040 MHz and developed the "Bethe logarithm," which we represent with the symbol $\beta_{n\ell}$ (beta, subscripts $n$ and $\ell$) [7]–[8]:

$$\begin{aligned}\beta_{1s} &= +2.984\ 128\ 555\ldots \\ \beta_{2s} &= +2.811\ 769\ 893\ldots \\ \beta_{2p} &= -0.030\ 016\ 708\ldots\end{aligned} \quad (9)$$

In 1948, Lamb and others calculated a relativistic value of 1052 MHz for the Lamb shift [9]–[10]. Richard Feynman introduced diagrams in 1949, to help visualize and keep track of the many contributions to this calculation and others [11]. Feynman's vertex function diagram (loosely resembling a lamb's head) is a main contributor to the more modern Lamb shift value of 1058 MHz:

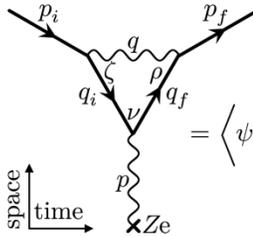

$$= \left\langle \psi(p_f) \left| \gamma^0 \left( \gamma^\nu F_1 + F_2 \frac{[\not{p},\gamma^\nu]}{4mc} \right) \right| \psi(p_i) \right\rangle \quad (10)$$

The vertex function (10) depicts possible changes in an electron's momentum (solid lines) when it interacts with electromagnetic waves (wavy lines) or photons. Its initial momentum can change from $p_i$ to $q_i$ when it emits a photon (momentum $q$), then from $q_i$ to $q_f$ via a Coulomb photon (momentum $p$), and from $q_f$ to $p_f$ when it absorbs its own photon of momentum $q$ again. In 1928, just after Gordon first solved the Dirac equation, he also decomposed it into form factors $F_1(p^2)$ and $F_2(p^2)$ [12]. Julian Schwinger calculated the second form factor for low momentum transfer $p^2 \to 0$, obtaining $F_2 = \alpha/(2\pi)$ in 1947 [13]. The first form factor $F_1$ is more complicated and wasn't found until 1948 [9]–[11]. By replacing the nucleus $Ze$ at "X" with a static Coulomb potential, we can reduce the commutator $[\not{p},\gamma^\nu] = \not{p}\,\gamma^\nu - \gamma^\nu\not{p}$ to matrix multiplication, or $-\boldsymbol{\gamma}\cdot\mathbf{p}$. Converting to position space leads to a 3D Dirac delta $\delta^3(\mathbf{r})$, which is commonly replaced by a Kronecker delta $\delta_{\ell 0}$ when sandwiched between large components $\psi^+$; these vanish except at the origin $\langle\psi^+|\delta^3(\mathbf{r})|\psi^+\rangle = |\psi^+(0)|^2 \delta_{\ell 0}$. Angular terms also reduce $\langle\psi^+|\boldsymbol{\sigma}\cdot\mathbf{L}/r^3|\psi^+\rangle = |\psi^+(0)|^2 c_{\ell j}$, where the angular expectation value is [14]–[17]:

$$c_{\ell j} = \frac{3m}{8\mu}\left(\frac{j(j+1) - \ell(\ell+1) - \frac{3}{4}}{(2\ell+1)\ell(\ell+1)} + \cdots\right)(1-\delta_{\ell 0}) \quad (11)$$

After further simplification, the form factors $F_1$ and $F_2$ become distributed among terms [14]–[17]:

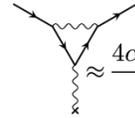

$$\approx \frac{4\alpha(Z\alpha)\hbar^3}{3m^2c}\overbrace{\left[\frac{1}{\pi a_0^3 n^3}\right]}^{|\psi^+(0)|^2}\left[\overbrace{\left(\ln\tau - \frac{3}{8} + \frac{3}{8}\right)}^{F_1}\overbrace{\phantom{\frac{3}{8}}}^{F_2}\delta_{\ell 0} + \overbrace{c_{\ell j}}^{F_2}\right] \quad (12)$$

The Lamb shift is an experimental observation, not tied to any single phenomenon, but to a multitude of effects. To calculate it with any accuracy, we must combine several diagrams like (12). We cannot determine the temporary quantity $\tau$ (tau) from $F_1$ alone, which leads to an "infrared catastrophe." Bremsstrahlung (braking radiation) suffers a catastrophe too, rendering both types of diagrams unusable on their own. However, when we combine these diagrams together, the catastrophes cancel each other out, and the logarithms merge into a dimensionless ratio of the electron rest energy $mc^2$ over the Hartree energy $\mu c^2 (Z\alpha)^2$:

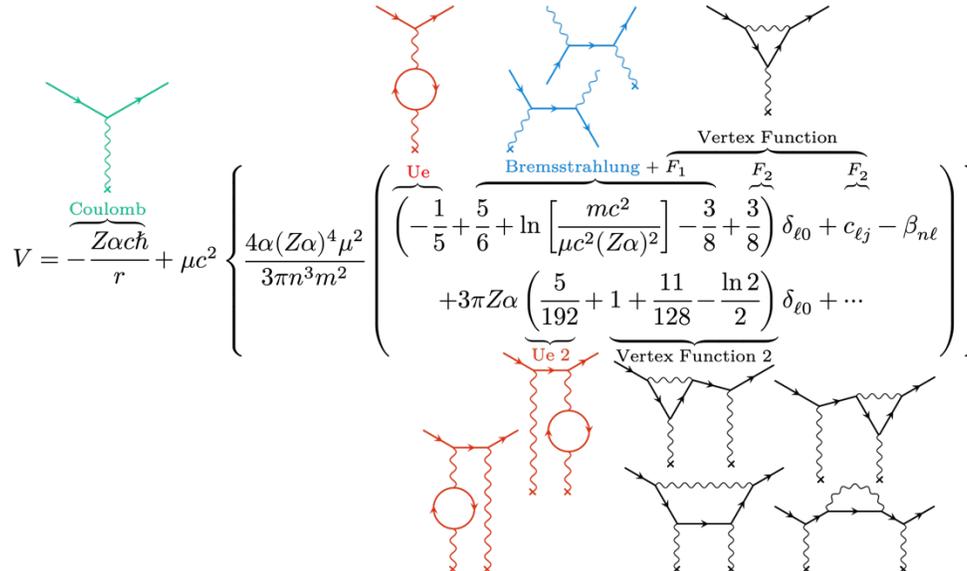

$$V = -\overbrace{\frac{Z\alpha c\hbar}{r}}^{\text{Coulomb}} + \mu c^2 \left\{\frac{4\alpha(Z\alpha)^4\mu^2}{3\pi n^3 m^2}\left(\begin{array}{c}\overbrace{\left(-\frac{1}{5}+\frac{5}{6}\right)}^{\text{Ue}} + \overbrace{\ln\left[\frac{mc^2}{\mu c^2(Z\alpha)^2}\right]}^{\text{Bremsstrahlung}+\overbrace{F_1}^{\text{Vertex Function}}} - \overbrace{\frac{3}{8}}^{F_2} + \overbrace{\frac{3}{8}}^{F_2}\right)\delta_{\ell 0} + c_{\ell j} - \beta_{n\ell} \\ +3\pi Z\alpha\left(\frac{5}{192}+1+\frac{11}{128}-\frac{\ln 2}{2}\right)\delta_{\ell 0} + \cdots\end{array}\right)\right\} \quad (13)$$

## 2. METHODS

Equation (13) depicts the Coulomb field centered at X exchanging a single photon with an electron to change its momentum (green). Sometimes however, a second photon escapes during transit to emerge as bremsstrahlung (blue). Other times, the electron reabsorbs this second photon before it can escape (vertex function). The photon can even split temporarily into a matter-antimatter pair in the case of the Uehling (Ue) potential. The Bethe logarithm $\beta_{n\ell}$ is the final contribution up to order $\alpha(Z\alpha)^4$ in the top line. The bottom line involves scenarios spanning two Coulomb photons (two X's) up to order $\alpha(Z\alpha)^5$, thereby generalizing the Uehling interaction (red) and vertex function (black). We sum all terms in the curly brackets { } to define a new dimensionless parameter lambda $\lambda$ for the Lamb shift:

$$\lambda = \frac{4\alpha(Z\alpha)^4 \mu^2}{3\pi n^3 m^2} \left[ \begin{pmatrix} \frac{19}{30} + \ln\left(\frac{m}{\mu(Z\alpha)^2}\right) \\ +3\pi Z\alpha\left(\frac{427}{384} - \frac{\ln 2}{2}\right) \\ +\cdots \end{pmatrix} \delta_{\ell 0} + c_{\ell j} - \beta_{n\ell} \right] \quad (14)$$

We calculate the dimensionless Lamb shift $\lambda$ (14) in the last column of TABLE I. The classic Lamb shift between the $2s_{\frac{1}{2}}$ and $2p_{\frac{1}{2}}$ states is $1044.815 - (-12.872) \approx 1058$ MHz. Multiplying $\lambda$ by $\mu c^2$, we combine it with Coulomb's law:

$$V = -\frac{Z\alpha c\hbar}{r} + \lambda \mu c^2 \quad (15)$$

The effective potential $V$ in (15) is identical to (13); it is simply written more concisely using $\lambda$ (14). Substituting this $V$ into the Dirac equation (6) yields:

$$\left(\frac{Z\alpha c\hbar}{r} \mp \mu c^2 + E - \lambda \mu c^2\right) R^\pm = c\hbar \left(\frac{k \mp 1}{r} \mp \frac{d}{dr}\right) R^\mp \quad (16)$$

We define dimensionless parameters $W^\mp$ and $w$ related to work or energy:

$$W^\mp = \sqrt{\frac{1 \mp w}{1 \pm w}} \qquad \text{where} \qquad w = \frac{E}{\mu c^2} - \lambda \quad (17)$$

Factoring out $\mu c^2$ on the left of the Dirac equation (16), then dividing both sides by $c\hbar$ gives:

$$\left(\frac{Z\alpha}{r} \mp \mu c \frac{1 \mp w}{\hbar}\right) R^\pm = \left(\frac{k \mp 1}{r} \mp \frac{d}{dr}\right) R^\mp \quad (18)$$

Multiplying $1 \mp w$ by a conjugate $1 = \sqrt{1 \pm w}/\sqrt{1 \pm w}$, and multiplying $\mu c/\hbar$ by $1 = (Z\alpha)/(Z\alpha)$, we find that the Bohr radius $a_0$ in (5) has naturally emerged as a scale of atomic length:

$$\left(\frac{Z\alpha}{r} \mp W^\mp \frac{\sqrt{1-w^2}}{a_0 Z\alpha}\right) R^\pm = \left(\frac{k \mp 1}{r} \mp \frac{d}{dr}\right) R^\mp \quad (19)$$

Only one denominator in (19) doesn't include $r$, so we define a dimensionless, independent variable $x$ in terms of another dimensionless quantity $v$:

$$x = \frac{2r}{a_0 v} \qquad \text{where} \qquad v = \frac{Z\alpha}{\sqrt{1-w^2}} \quad (20)$$

Multiplying (19) by $a_0 v/2$ and replacing denominators of $r$ with $x$ completes the nondimensionalization process:

$$\left(\frac{Z\alpha}{x} \mp \frac{W^\mp}{2}\right) R^\pm = \mp \left(\frac{1 \mp k}{x} + \frac{d}{dx}\right) R^\mp \quad (21)$$

Now to solve the dimensionless Dirac equation (21), we consider decay exponentials $e^{-bx}$, which converge for large $x$ and positive $b$. Polynomials $x^{u/2-1}$ will converge for small $x$ and $u \gtrsim 2$. Like Pidduck, we also want to take advantage of the recurrence relations for the Laguerre polynomials $L_g^u(x)$, for example [4]:

$$x \frac{d}{dx} L_g^u = g L_g^u - (g+u) L_{g-1}^u \quad (22)$$

With these considerations, we seek new solutions with normalization factors $N^\pm$ and coefficients $b$, $u$, and $C^\pm$:

$$R^\pm = N^\pm e^{-bx} x^{u/2-1} (C^\pm y + xy') \quad (23)$$

Using a temporary quotient $C^\pm = P^\pm/Q^\pm$, we multiply our trial solutions (23) by $Q^\pm/N^\pm$:

$$\frac{Q^\pm}{N^\pm} R^\pm = e^{-bx} x^{u/2-1} (P^\pm y + Q^\pm xy') \quad (24)$$

Substituting (24) into the Dirac equation (21), we immediately divide by $e^{-bx} x^{u/2-1}$ and use primes to denote derivatives (e.g., $y' = dy/dx$ and $y'' = d^2y/dx^2$):

$$\left(\frac{Z\alpha}{x} \mp \frac{W^\mp}{2}\right)(P^\pm y + Q^\pm xy')$$
$$= \mp \left( \begin{matrix} (P^\mp y + Q^\mp xy')\left(\frac{1 \mp k}{x} - b + \frac{u/2-1}{x}\right) \\ +P^\mp y' + Q^\mp y' + Q^\mp xy'' \end{matrix} \right) \quad (25)$$

TABLE I. Lamb shift contributions in MHz, and dimensionless equivalents.

| State | Order $\alpha(Z\alpha)^4$ in MHz | | | | | $\alpha(Z\alpha)^5$ in MHz | | MHz | Dimensionless |
| --- | --- | --- | --- | --- | --- | --- | --- | --- | --- |
| $n\ell j$ | Uehling | Brem + $F_1$ | $F_2\,\delta_{\ell 0}$ | $F_2\,c_{\ell j}$ | $-\beta_{n\ell}$ | Ue 2 | Vert 2 | Total | Total $\lambda$ (14) |
| $1s_{\frac{1}{2}}$ | $-216.675$ | $11\,158.118$ | $406.267$ | | $-3\,232.943$ | $1.940$ | $55.090$ | $8\,171.797$ | $66\,180 \times 10^{-15}$ |
| $2s_{\frac{1}{2}}$ | $-27.084$ | $1\,394.765$ | $50.783$ | | $-380.777$ | $0.243$ | $6.886$ | $1\,044.815$ | $8\,461 \times 10^{-15}$ |
| $2p_{\frac{1}{2}}$ | | | | $-16.937$ | $4.064$ | | | $-12.872$ | $-104 \times 10^{-15}$ |
| $2p_{\frac{3}{2}}$ | | | | $8.468$ | $4.064$ | | | $12.533$ | $101 \times 10^{-15}$ |

To utilize the Laguerre polynomials, we will rewrite the Dirac equation (25) as a Laguerre equation:

$$0 = xy'' + (u+1-x)y' + gy \qquad (26)$$

Subtracting the left side of (25) over to the right and rearranging terms in orders of $x$ and $y$, we divide by $\mp Q^\mp$ to better match the Laguerre equation (26):

$$0 = xy''$$
$$+ \left[ \overbrace{\left( \frac{u}{2} \mp k \pm Z\alpha \frac{Q^\pm}{Q^\mp} + C^\mp \right)}^{u} + \overbrace{1}^{1} - x \left( \frac{W^\mp Q^\pm}{2Q^\mp} + b \right) \right] y'$$
$$+ \left[ \underbrace{\left( \frac{u}{2} \mp k \pm Z\alpha \frac{P^\pm}{P^\mp} \right)}_{0} \frac{1}{x} C^\mp - \underbrace{\left( \frac{W^\mp Q^\pm}{2Q^\mp} C^\pm + bC^\mp \right)}_{-g} \right] y \qquad (27)$$

The lowest-order terms, parenthesized in the bottom left of (27), don't appear in the Laguerre equation (26), so we set them equal to zero, add $\mp Z\alpha P^\pm/P^\mp$ to both sides, then divide by $\mp Z\alpha$ to obtain a ratio of coefficients:

$$\frac{P^\pm}{P^\mp} = \frac{k \mp u/2}{Z\alpha} \qquad (28)$$

There are two ways to solve for the ratio of $P^+$ to $P^-$: using the top signs (first equation) or flipping the equation upside down for the bottom signs (second equation). We set both equations equal, cross multiply, and solve for $u$:

$$u = 2\sqrt{k^2 - (Z\alpha)^2} \qquad (29)$$

The last parenthesis in the middle row of (27) should equal one, so we set:

$$b = \frac{1}{2} \qquad \text{and} \qquad \frac{Q^\pm}{Q^\mp} = W^\pm \qquad (30)$$

We set the first parenthesis in (27) equal to $u$ as labeled, substitute (30), solve for $C^\mp = u/2 \pm k \mp Z\alpha W^\pm$, multiply $Z\alpha W^\pm$ by the conjugate $1 = \sqrt{1 \pm w}/\sqrt{1 \pm w}$, and rewrite the last term utilizing $v$ in (20):

$$C^\mp = \frac{u}{2} \pm k \mp v - vw \qquad (31)$$

Substituting (30) and (31) into the last remaining terms of (27) yields:

$$-g = \frac{W^\mp}{2} \overbrace{\widetilde{W^\pm}}^{\frac{Q^\pm}{Q^\mp}} \overbrace{\left[ \frac{u}{2} \mp k \pm v - vw \right]}^{C^\pm = \frac{P^\pm}{Q^\pm}} + \overbrace{\frac{1}{2}}^{b} \overbrace{\left[ \frac{u}{2} \pm k \mp v - vw \right]}^{C^\mp = \frac{P^\mp}{Q^\mp}} \qquad (32)$$

From (17) we can see that $W^\mp W^\pm = 1$, so several terms in (32) cancel:

$$g = vw - \frac{u}{2} \qquad (33)$$

Substituting (22) and (30)–(33) into our trial solutions (23), we use Laguerre polynomials to replace $y \to L_g^u(x)$:

$$R^\pm = N^\pm e^{-\frac{bx}{x/2}} x^{u/2-1}$$
$$\times \left[ \underbrace{\left( \frac{u}{2} \mp k \pm v - vw + \underbrace{vw - \frac{u}{2}}_{g} \right) L_g^u}_{C^\pm} - (g+u) L_{g-1}^u \right] \qquad (34)$$

Additionally substituting $v$ from (20) into (33), we solve for $w$:

$$w = \frac{1}{\sqrt{1 + \left( \frac{Z\alpha}{g + u/2} \right)^2}} \qquad (35)$$

Substituting (35) back into (20) reveals that $v$ is an "apparent" principal quantum number, sometimes taking the values of principal quantum number $n$ [18]:

$$v = \sqrt{(g + u/2)^2 + (Z\alpha)^2} \qquad (36)$$

## 3. RESULTS

We observe that $w$, which contains the Lamb shift $\lambda$ in (17), cancels out in (34). Values like $v$ (36) can be expressed independently of $\lambda$, so $\lambda$ vanishes from our eigenfunctions:

$$R^\pm = N^\pm \sqrt{e^{-x} x^{u-2}} [\pm(v-k)L_g^u(x) - (g+u)L_{g-1}^u(x)] \qquad (37)$$

To ensure the last Laguerre polynomial in (37) remains specified whenever $g = 0$, we include a special definition for $L_{-1}^u(x) = 0$ in TABLE II. In 1930, Karl Bechert integrated confluent hypergeometric functions to find both large $N^+$ and small $N^-$ normalization factors, which we can write with gamma functions $\Gamma(z)$ or factorials $(z-1)!$ [19]:

$$N^\pm = \sqrt{\frac{2(1 \pm w)\Gamma(1+g)}{a_0^3 v^4 (v-k) \Gamma(1+g+u)}} \qquad (38)$$

If we neglect spin in the limit $k \to \ell$, then $g \to n - \ell$. If we also neglect the fine structure $\alpha \to 0$, then according to TABLE III: $u \to 2\ell$, $v \to n$, and $w \to 1$, canceling the small normalization factor $N^-$ with $1 - w$ in (38), so only large components $R^+$ and $N^+$ remain in (37)–(38). Furthermore, we can apply the recurrence relation (22) in reverse to combine both Laguerre polynomials back into a single derivative, then employ $(d/dx)L_{n-\ell}^{2\ell} = -L_{n-\ell-1}^{2\ell+1}$:

$$\lim_{\substack{k \to \ell \\ \alpha \to 0}} (R^\pm) = -\underbrace{\sqrt{\frac{4(n-\ell-1)!}{a_0^3 n^4 (n+\ell)!}}}_{N^+} \sqrt{e^{-x}} x^\ell L_{n-\ell-1}^{2\ell+1}(x) \qquad (39)$$

Erwin Schrodinger obtained (39) in 1926 [20]. We now plot the radial probability distribution for the Dirac states $1s_{\frac{1}{2}}$, $2s_{\frac{1}{2}}$, $2p_{\frac{1}{2}}$, and $2p_{\frac{3}{2}}$ from (37)–(38), and Schrodinger states 1s, 2s, and 2p from (39) in FIG 1.

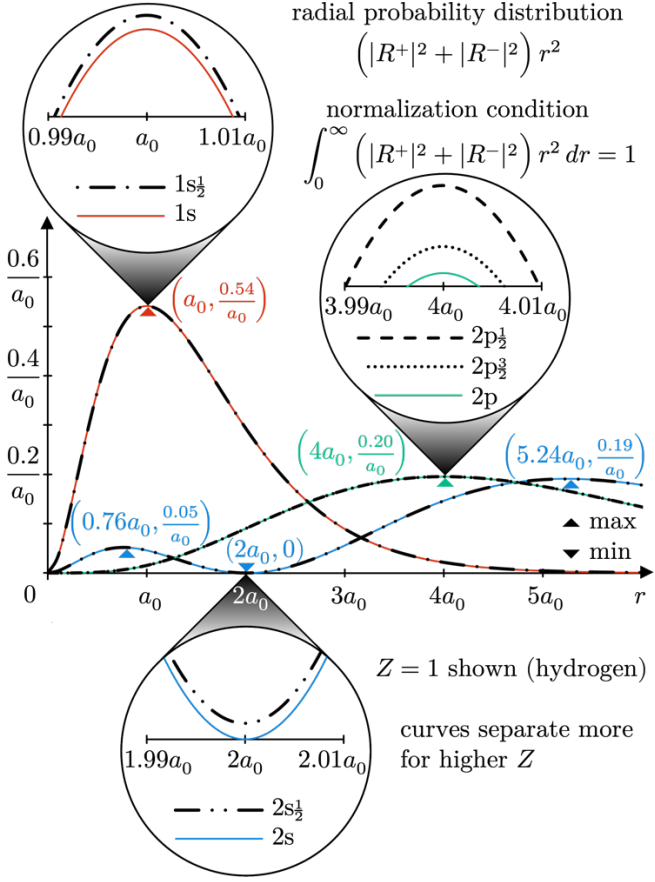

FIG 1. The Schrodinger ground state 1s (solid red curve) has a maximum peak, one Bohr radius $a_0$ from the origin $(a_0, \ 0.54/a_0)$. The Dirac ground state $1s_{\frac{1}{2}}$ (dash-dotted) peak is raised up slightly to the left at $r = a_0\sqrt{1-(Z\alpha)^2}$. The Schrodinger 2s state (solid blue) node is at $(2a_0, 0)$. The Dirac $2s_{\frac{1}{2}}$ state (dash-double-dotted) lacks this node, rising slightly to the left again. The Schrodinger 2p state (solid green) peaks at $(4a_0, \ 0.20/a_0)$. The Dirac $2p_{\frac{3}{2}}$ state (dotted curve) is raised slightly left at $r = 2a_0\sqrt{4-(Z\alpha)^2}$ and the $2p_{\frac{1}{2}}$ state (dashed curve) rises even further up and further left (note that each leftward shift is too small to see clearly at this scale).

TABLE II. Low-order Laguerre polynomials.

| | |
|---|---|
| $L^u_{-1}(x) = 0$ | (non-standard definition) |
| $L^u_0(x) = 1$ | |
| $L^u_1(x) = [(u+1) - x]$ | |
| $L^u_2(x) = [(u+1)(u+2) - 2(u+2)x + x^2]/2!$ | |

We can substitute $g$ and $u$ into $w$ from TABLE III and set this equal to our definition of $w = E/\mu c^2 - \lambda$ in (17), then add $\lambda$ to both sides. Rest mass was not included in this definition, so after multiplying by $\mu c^2$, we subtract another $\mu c^2$ from the right side to obtain our final energy levels:

$$E = \mu c^2 \left( \frac{1}{\sqrt{1 + \left(\frac{Z\alpha}{n - |k| + \sqrt{k^2 - (Z\alpha)^2}}\right)^2}} + \lambda - 1 \right) \quad (40)$$

In the last column of TABLE IV, we calculate these energy levels (40) in electronvolts (eV).

## 4. DISCUSSION

Upon placing the Lamb shift $\lambda$ into the Dirac equation, we found it cancels out in the radial functions (37)–(38), likely because it lacks explicit $r$ dependence (e.g., recall the common replacement $\langle \psi^+ | \delta^3(\mathbf{r}) | \psi^+ \rangle = |\psi^+(0)|^2 \delta_{\ell 0}$). Hence, it is challenging to demonstrate specific effects of the Lamb shift upon probability distributions. Since $a_0$ and $\lambda$ are tiny numbers less than $10^{-10}$, multiplying them produces yet smaller numbers, making the Lamb shift imperceptible in the magnified portions of FIG 1. Hyperfine splitting is even smaller for low $Z$. Using $\pm f$ to denote the Lamb shift $\lambda$ plus hyperfine shifts, we could replace $x \to 2r/[a_0(v \pm f)]$ in (37) and $v^4 \to v(v \pm f)^3$ in (38) to illustrate these shifts. But in the limit as $f \to 0$, our radial solutions (37)–(38) are fundamentally the Dirac eigenfunctions obtained in 1928, and revert to the Schrodinger wavefunctions from 1926 if the limits in (39) are applied [2]–[4], [20]. As $r \to 0$, the large, Schrodinger functions squared $|R^+(0)|^2$ further connect back to (12) via the probabilities $|\psi^+(0)|^2$.

The last column in TABLE IV demonstrates that our energy levels (40) replicate the classic Lamb shift for the $2s_{\frac{1}{2}}$ and $2p_{\frac{1}{2}}$ states; dividing by Planck's constant confirms this: $[-3.399\ 624\ 069 - (-3.399\ 628\ 443)]/h \approx 1058$ MHz. Finally, in the limit as $\lambda \to 0$, our nondegenerate energy levels (40) revert to the traditional, degenerate energy levels obtained by Arnold Sommerfeld and Wilhelm Lenz in 1916 [21].

TABLE III. Summary of (7), (8), (20), (29), (35), and (36).

| | |
|---|---|
| $k = \pm(j + \frac{1}{2})$ when $j = \ell \mp \frac{1}{2}$ | $w = \dfrac{1}{\sqrt{1 + \left(\dfrac{Z\alpha}{g + u/2}\right)^2}}$ |
| $g = n - |k|$ | |
| $u = 2\sqrt{k^2 - (Z\alpha)^2}$ | |
| $v = \sqrt{(g + u/2)^2 + (Z\alpha)^2}$ | $x = 2r/(a_0 v)$ |

TABLE IV. Quantities for specific low-order states.

| State | Quantum Numbers | | | | | Dimensionless Parameters | | | | Energy Level |
|---|---|---|---|---|---|---|---|---|---|---|
| $n\ell j$ | $n$ | $\ell$ | $j$ | $k$ | $g$ | $u$ | $v$ | $w$ | $x$ | $E$ (40) in eV |
| $1s_{\frac{1}{2}}$ | 1 | 0 | 1/2 | $-1$ | 0 | $2\sqrt{1-(Z\alpha)^2}$ | 1 | $u/2$ | $2r/a_0$ | $-13.598\ 434\ 504$ |
| $2s_{\frac{1}{2}}$ | 2 | 0 | 1/2 | $-1$ | 1 | $2\sqrt{1-(Z\alpha)^2}$ | $\sqrt{u+2}$ | $v/2$ | $2r/(a_0 v)$ | $-3.399\ 624\ 069$ |
| $2p_{\frac{1}{2}}$ | 2 | 1 | 1/2 | $+1$ | 1 | $2\sqrt{1-(Z\alpha)^2}$ | $\sqrt{u+2}$ | $v/2$ | $2r/(a_0 v)$ | $-3.399\ 628\ 443$ |
| $2p_{\frac{3}{2}}$ | 2 | 1 | 3/2 | $-2$ | 0 | $2\sqrt{4-(Z\alpha)^2}$ | 2 | $u/4$ | $r/a_0$ | $-3.399\ 583\ 078$ |